\documentclass[pdftex,twocolumn,epjc3]{svjour3}
\usepackage{amsmath}  
\usepackage{amsfonts} 
\usepackage{graphicx}
\usepackage{subeqnarray}
\usepackage{color}
\usepackage[normalem]{ulem}
\usepackage{url}
\usepackage{multirow}
\usepackage{paralist}
\RequirePackage[T1]{fontenc}

\smartqed  % flush right qed marks, e.g. at end of proof

\RequirePackage{graphicx}
\RequirePackage{mathptmx}      % use Times fonts if available on your TeX system
\RequirePackage{flushend}
\RequirePackage[numbers,sort&compress]{natbib}
\RequirePackage[colorlinks,citecolor=blue,urlcolor=blue,linkcolor=blue]{hyperref}

\journalname{Eur. Phys. J. C}
\hypersetup{colorlinks=false}
\hypersetup{linktocpage}
\definecolor{darkblue}{RGB}{16,78,139}
\newcommand{\beq}{\begin{equation}}
\newcommand{\eeq}{\end{equation}}
\newcommand{\beqn}{\begin{eqnarray}}
\newcommand{\eeqn}{\end{eqnarray}}
\newcommand{\beqs}{\begin{subeqnarray}}
\newcommand{\eeqs}{\end{subeqnarray}}
\newcommand{\nn}{\nonumber}

\begin{document}

%\title{Tidal Forces in Reissner-Nordstr\"om Spacetimes}

%\author{Lu\'is C. B. Crispino}
%\email{crispino@ufpa.br}
%\affiliation{Faculdade de F\'{\i}sica, Universidade 
%Federal do Par\'a, 66075-110, Bel\'em, Par\'a, Brazil.}

%\author{Atsushi Higuchi}
%\email{atsushi.higuchi@york.ac.uk}
%\affiliation{Department of Mathematics, University of York, YO10 5DD Heslington, York, United Kingdom.}

%\author{Leandro A. Oliveira}
%\email{leandro.oliveira@york.ac.uk}
%\affiliation{Faculdade de F\'{\i}sica, Universidade 
%Federal do Par\'a, 66075-110, Bel\'em, Par\'a, Brazil.}
%\affiliation{Department of Mathematics, University of York, YO10 5DD Heslington, York, United Kingdom.}

%\author{Ednilton S. de Oliveira}
%\email{ednilton@pq.cnpq.br}
%\affiliation{Faculdade de F\'{\i}sica, Universidade 
%Federal do Par\'a, 66075-110, Bel\'em, Par\'a, Brazil.}

%\date{\today}

\title{Tidal Forces in Reissner-Nordstr\"om Spacetimes}


\author{Lu\'is C. B. Crispino\thanksref{e1,addr1}
        \and
        Atsushi Higuchi\thanksref{e2,addr2}
        \and
        Leandro A. Oliveira\thanksref{e3,addr2}
        \and
        Ednilton S. de Oliveira\thanksref{e4,addr1}
}

%\thankstext[$\star$]{t1}{Thanks to the title}
\thankstext{e1}{e-mail: crispino@ufpa.br}
\thankstext{e2}{e-mail: atsushi.higuchi@york.ac.uk}
\thankstext{e3}{e-mail: leandro.oliveira@york.ac.uk}
\thankstext{e4}{e-mail: ednilton@pq.cnpq.br}

\institute{Faculdade de F\'{\i}sica, Universidade Federal do Par\'a, 66075-110, Bel\'em, Par\'a, Brazil\label{addr1}
          \and
          Department of Mathematics, University of York, YO10 5DD Heslington, York, United Kingdom\label{addr2}
}

\date{Received: date / Accepted: date}

\maketitle

\begin{abstract}
%%%%%%%%%%%%%%%%%%%%%%%%%%%%%%%%%%%%%%%
We analyze the tidal forces produced in the spacetime
of Reissner-Nordstr\"om black holes. We point out that the radial component of the tidal force changes sign just outside the event horizon
if the charge-to-mass ratio is close to $1$ unlike in Schwarzschild spacetime of uncharged black holes, and that the angular
component changes sign between the outer and inner horizons.  We solve the
geodesic deviation equations for radially falling bodies towards the charged black hole.  We 
find, for example, that the radial component of the geodesic deviation vector starts decreasing inside the event horizon unlike in the Schwarzschild case.
\end{abstract}

\maketitle

%%%%%%%%%%%%%%%%%%%%%%%%%%%%%%%%%%%%%%%
\section{Introduction}
%%%%%%%%%%%%%%%%%%%%%%%%%%%%%%%%%%%%%%%
Black holes are objects of great fascination for the scientific community
as well as for the general public, especially because of their remarkable physical properties.
Schwarzschild (uncharged) black holes -- the simplest case -- have been
extensively investigated over the years.
However, less attention has been given
to Reissner-Nordstr\"om (electrically charged) and Kerr (rotating)
black holes. The importance of studying these more complex black
holes lies in the fact that they present a new set of phenomena that are
not present in Schwarzschild spacetimes. Examples are the Penrose
process~\cite{nps229_177:Penrose&Floyd}, superradiance~\cite{jetp37_28},
and interconversion between spin 1 and 2 fields~\cite{prd14_3274, PhysRevD.80.104026}, as well as electromagnetic helicity-reversing processes~\cite{PhysRevD.90.064027}.
Reissner-Nordstr\"om black holes are of special interest because
they allow one to analyze extreme spacetime configurations with spherical
symmetry. For example, it has recently been found that extreme
Reissner-Nordstr\"om black holes absorb and scatter gravitational and
electromagnetic waves equally~\cite{prd84_084048, cdho2}. This equality is a
consequence of the supersymmetry
that relates photons and gravitons in the extreme Reissner-Nordstr\"om spacetime~\cite{prd55_R4529}.

In this paper we focus on Reissner-Nordstr\"om black holes,
which are spherically symmetric, have non-zero electric charge
but no angular momentum.
They are exact solutions of the Einstein-Maxwell
equations~\cite{Chand},
and in the case of vanishing electric charge, they reduce to the
Schwarzschild black holes.

It is well known that a body falling
towards the event horizon of a static uncharged black
 hole experiences stretching in the radial direction and compression
 in the angular directions~\cite{DInverno2, Lasenby, Schutz, MTW, Hartle}.
 However, whether a body may experience stretching or compression
in either direction (radial or angular) in Reissner-Nordstr\"om spacetime depends on the charge-to-mass ratio of the black hole and where the
body is located (see, e.g.\ Ref.~ \cite{artigo2}).
At certain points of Reissner-Nordstr\"om spacetimes, the tidal forces in the radial or angular direction change their sign unlike in Schwarzschild spacetime.  In this paper we describe the tidal forces in Reissner-Nordstr\"om spacetime in detail.  We then
solve the geodesic deviation equations to analyze the changes in
 size of a test body consisting of neutral dust particles infalling
radially towards the Reissner-Nordstr\"om black hole.  
We also point out that the tidal forces can be understood within Newtonian Mechanics if an extra force coming
from General Relativity is added, while the
geodesic deviation needs to be analyzed using full General Relativity.
The remainder of this paper is organized as follows.
In Sec.~\ref{sectionV} we briefly review relevant facts about Reissner-Nordstr\"om black holes.
We analyze geodesics in Reissner-Nordstr\"om spacetime in Sec.~\ref{sectionII}
 and study tidal forces for charged static black holes in Sec.~\ref{sectionIII}.
In Sec.~\ref{SecVI} we obtain the solutions of the geodesic deviation equations.
Then we present our conclusion in Sec.~\ref{sectionIV}. We use the metric signature $(+,-,-,-)$ and set the speed of light $c$ and Newtonian
gravitational constant $G$ to $1$ throughout this paper.

%%%%%%%%%%%%%%%%%%%%%%%%%%%%%%%%%%%%%%%
\section{Reissner-Nordstr\"om Black Holes \label{sectionV}}
%%%%%%%%%%%%%%%%%%%%%%%%%%%%%%%%%%%%%%%
The line element of a static charged black hole is given by~\cite{Chand}
\beqn
d{s}^2 &=&g_{\mu\nu}dx^{\mu}dx^{\nu}\nn\\
&=& f(r)dt^2 - f(r)^{-1}dr^2 -
r^2(d\theta^2 + \sin^2\theta d\phi^2),\nn\\
\label{general-eq}
\eeqn
with
\beq
f(r) =  1-\frac{2M}{r} +
\frac{q^2}{r^2}, \label{line}
\eeq
where $M$ and $q$ are the mass and charge (in gaussian units) of the
black hole, respectively.

There are three possible configurations
(with $q\neq0$) for the Reissner-Nordstr\"om spacetime:
\begin{inparaenum}[(i)]\\
 \item{For $q^2/M^2< 1$,
we have a Reissner-Nordstr\"om black hole with two horizons.}\\
\item{For $q^2/M^2 = 1$,
we have an extremely charged Reissner-Nordstr\"om black
hole, with the event horizon located at
$r_{+}=r_{-}=M$.}\\
\item{For $q^2/M^2 >  1$,
we have a naked singularity.}
\end{inparaenum}
Here we will only consider the cases in which
$q^2/M^2 \leq 1$ \  (black hole spacetimes).
(Naked singularities do not occur in nature if the cosmic censor conjecture~\cite{PenroseCS} is true.)
The radial coordinates of the horizons, i.e. the zero(s) of the function $f(r)$, are
\beqn
r_{+} & = & M+
\sqrt{M^2- q^2},
\label{rh+}\\
r_{-} & = & M -
\sqrt{M^2- q^2}.
\label{rh-}
\eeqn
Eq.~(\ref{rh+}) gives the location of the external
horizon, or the event horizon, of the black hole and
Eq.~(\ref{rh-}) gives the location of the internal
horizon, or the Cauchy horizon, of the black hole~\cite{artigo7}.

%%%%%%%%%%%%%%%%%%%%%%%%%%%%%%%%%%%%%%%
\section{Radial Geodesics in Reissner-Nordstr\"om Spacetimes}
 \label{sectionII}
%%%%%%%%%%%%%%%%%%%%%%%%%%%%%%%%%%%%%%%
We consider a radial geodesic motion 
in spherically symmetric spacetimes with line element \eqref{general-eq}.
By letting $ds=d\tau$ in Eq.~\eqref{general-eq} we obtain~\cite{Wald}
\beq
f(r)\dot{t}^2 - f(r)^{-1}\dot{r}^2 = 1,
\label{line-2}
\eeq
where the dot represents the differentiation with respect to the proper time $\tau$.  We have let $\dot{\theta} = \dot{\phi} = 0$ because
the motion is radial by assumption.
As is well known,  $E=f(r)\dot{t}$ is conserved and is interpreted as the mechanical energy per unit mass
of the particle. Substituting this equation
into Eq.~(\ref{line-2}), we obtain
\beq
\frac{\dot{r}^2}{2} = \frac{E^2-f(r)}{2}. \label{dotr-squared}
\eeq
For the radial infall of a test particle from rest at position $b$, we obtain 
$E=\sqrt{f(r=b)}$
from Eq.~(\ref{dotr-squared})~\cite{Martel}.

By defining the ``Newtonian radial acceleration"~\cite{Symon} as
\beq
A^{(R)}  \equiv \ddot{r},
\label{equiv}
\eeq 
we find from Eq.~\eqref{dotr-squared} that
\beq
A^{(R)} = -\frac{f'(r)}{2},
\eeq
where the prime denotes the differentiation with respect to
the radial coordinate $r$.
For Reissner-Nordstr\"om spacetime this becomes
\beq
A^{(R)} = -\frac{M}{r^2} + \frac{q^2}{r^3}.
\label{freefall-3}
\eeq
This gives us the ``Newtonian radial acceleration"
that Reissner-Nordstr\"om
black hole ``exerts'' on a neutral freely falling massive test body.%,

The term $q^2/r^3$ in Eq.~(\ref{freefall-3}) represents a
purely relativistic effect.
It has been analyzed in the literature (see Ref.~ \cite{artigo6}), and has
been a theme of discussion (see Refs.~\cite{artigo4}
and~ \cite{artigo5}).
 It is interesting to note that
the test particle falling freely from rest at $r=b> r_{+}$ (for $q\neq 0$)
would bounce back at $R^{\text{stop}}$. The radius $R^{\text{stop}}$ can readily be found as a root of $E^2 - f(r)$ as
\beq
R^{\text{stop}} = \frac{bq^2}{2Mb-q^2 },
\label{Rstop}
\eeq
where we recall that $b$ is the initial position of test particle (starting from rest). The radius
$R^{\text{stop}}$ is always located inside the internal (Cauchy) horizon. In the limit $b\rightarrow \infty$, 
one finds $R^{\text{stop}} \rightarrow q^2/2M$. In the maximal analytic extension of Reissner-Nordstr\"om spacetime 
the particle, after bouncing back at $R^{\text{stop}}$, would emerge in a different asymptotically flat region of the spacetime~\cite{Lasenby}.  
For more details about the
physics of freely falling neutral particles in
Reissner-Nordstr\"om black holes, we refer the
reader to Refs.~\cite{DInverno2} and~\cite{Lasenby}. We note in passing that the Cauchy horizon is known to be unstable~\cite{Simpson-Penrose}.  Thus, the part of the spacetime
beyond the Cauchy horizon in the maximal analytic extension of Reissner-Nordstr\"om spacetime is unphysical.

%%%%%%%%%%%%%%%%%%%%%%%%%%%%%%%%%%%%%%%
\section{Tidal Forces in Reissner-Nordstr\"om Spacetime on a Neutral Body in Radial Free Fall \label{sectionIII}}
%%%%%%%%%%%%%%%%%%%%%%%%%%%%%%%%%%%%%%%
Now let us turn our attention to the tidal forces acting in Reissner-Nordstr\"om spacetime.
As is well known~\cite{DInverno2,Lasenby},
the equation for the spacelike components of
the geodesic deviation vector $\eta^{\mu}$ that describes the distance between two infinitesimally close particles in free fall is given by
\beq
\frac{D^2\eta^\mu}{D\tau^2} -
R^{\mu}_{\ \sigma \nu \rho}v^{\sigma}v^{\nu}\eta^\rho=0,
\label{Deviation2}
\eeq
where $v^\nu$ is the unit vector tangent to the geodesic.
We introduce the tetrad basis for radial free-fall reference frames:
\beqn
&&\hat{e}^{\ \mu}_{\widehat{0}} = \left( \frac{E}{f},\,-\sqrt{E^2-f},\,0,\,0\right) , \nn\\
\nn\\
&&\hat{e}^{\ \mu}_{\widehat{1}} = \left( \frac{-\sqrt{E^2-f}}{f},\,E,\,0,\,0\right) , \nn\\
\nn\\
&&\hat{e}^{\ \mu}_{\widehat{2}} = r^{-1} \left( 0,\,0,\,1,\,0\right) , \nn\\
\nn\\
&&\hat{e}^{\ \mu}_{\widehat{3}} = \left( r\sin{\theta}\right) ^{-1} \left( 0,\,0,\,0,\,1\right),\nn
\eeqn
where $(x^0,x^1,x^2,x^3) = (t,r,\theta,\phi)$.
These unit vectors satisfy 
the following orthonormality condition:
\beq
\hat{e}^{\ \alpha}_{\widehat{\mu}}\hat{e}_{\widehat{\nu} \alpha} = \eta_{\widehat{\mu}\widehat{\nu}},
\eeq
where $\eta_{\widehat{\mu}\widehat{\nu}}$ is the Minkowski metric (for more details, see Ref.~\cite{DInverno2}).
We have that $\hat{e}^{\ \mu}_{\widehat{0}} = v^\mu$.
The geodesic deviation vector, also called the Jacobi vector, can be expanded as
\beq
\eta^{\mu} = \hat{e}_{\widehat{\nu}}^{\ \mu}\eta^{\widehat{\nu}}.
\eeq
Here we note that $\eta^{\widehat{0}} =0$~\cite{DInverno2}.

The non-vanishing independent components of the Riemann tensor
in spherically symmetric spacetimes, including the Reissner-Nordstr\"om spacetime, are (see, e.g., Ref.~ \cite{Wald})
\beqn
&&R^1_{\ 212}= -\frac{rf^{\prime}}{2},\nn\quad\quad
R^1_{\ 313}= -\frac{rf^{\prime}}{2}\sin^2\theta,\nn\quad\quad
\nn\\
&&R^1_{\ 010}= \frac{ff^{\prime \prime}}{2},\nn\quad\quad
R^2_{\ 323}= \left(1-f\right)\sin^2\theta,\nn\\
\nn\\
&&R^2_{\ 020}= -\frac{ff^{\prime}}{2r},\nn\quad\quad
R^3_{\ 030}= -\frac{ff^{\prime}}{2r}.\nn
\eeqn
Using these expressions in Eq.~\eqref{Deviation2} and noting that the vectors 
${\hat{e}_{\widehat{\nu}}}^{\ \mu}$ 
are all parallelly transported along the geodesic, 
we find 
the following equations for tidal forces in radial free-fall reference frames (see, e.g., Ref.~\cite{artigo2}):
\beqn
\ddot{\eta}^{\widehat{1}} & = & -\frac{f^{\prime \prime}}{2}\eta^{\widehat{1}}, \label{radials}\\ 
\ddot{\eta}^{\widehat{i}} & = & -\frac{f^{\prime}}{2r}\eta^{\widehat{i}}, \label{theta}% \\
\eeqn
where $i=2,\,3$.

Substituting the explicit form~\eqref{line} of $f(r)$ in Reissner-Nordstr\"om spacetime into Eqs.~(\ref{radials}) and~(\ref{theta}) we see that
the tidal forces in this spacetime depend on the
mass and charge of a black hole.
We also see that the radial and angular tidal forces may
vanish, in contrast to what happens in the
Schwarzschild spacetime ($q=0$)~\cite{DInverno2, Lasenby, Schutz, MTW}.
We note that the expressions of the tidal forces, given by Eqs.~(\ref{radials}) and~(\ref{theta}),
are identical to the Newtonian tidal forces with the force $-f'/2$ in the radial direction.
In the remainder of this paper we study Eqs.~(\ref{radials}) and (\ref{theta}) for Reissner-Nordstr\"om spacetime in detail.

%%%%%%%%%%%%%%%%%%%%%%%%%%%%%%%%%%%%%%%
\subsection{Radial Tidal Force}
%%%%%%%%%%%%%%%%%%%%%%%%%%%%%%%%%%%%%%%
From Eqs.~\eqref{line} and \eqref{radials} it can readily be shown that 
the radial tidal force vanishes at $r=R^{\text{rtf}}_0$, where
\beq
R^{\text{rtf}}_0 = \frac{3q^2}{2M}.
\label{rtf}
\eeq
By comparing this and the expression~\eqref{rh+} for the event horizon we find that,
if $2\sqrt{2}/3 \leq q/M \leq 1$~\footnote{Since we are dealing with neutral test particles, the results presented here are valid for both negatively and positively charged black holes. That is, although we are restricting the analysis to positively charged black holes, all results presented here depend only on the magnitude of the black hole charge and apply to negatively charged black holes as well.},
the radial tidal
 force inverts its direction and becomes compressing just outside
the event horizon.
The radial tidal force takes a maximum value at $R^{\text{rtf}}_{\text{max}}$ where
\beq
R^{\text{rtf}}_{\text{max}} = \frac{2q^2}{M}.
 \label{rmax}
\eeq
The maximum radial stretching at $r = R^{\text{rft}}_{\text{max}}$ is~\cite{artigo6}
\beq
\ddot{\eta}^{\widehat{1}}|_{\text{max}}=
\frac{M^4}{16q^6} \eta^{\widehat{1}}.
\eeq

\begin{figure}[htpb!]
	\centering
		\includegraphics[width=1.0\columnwidth]{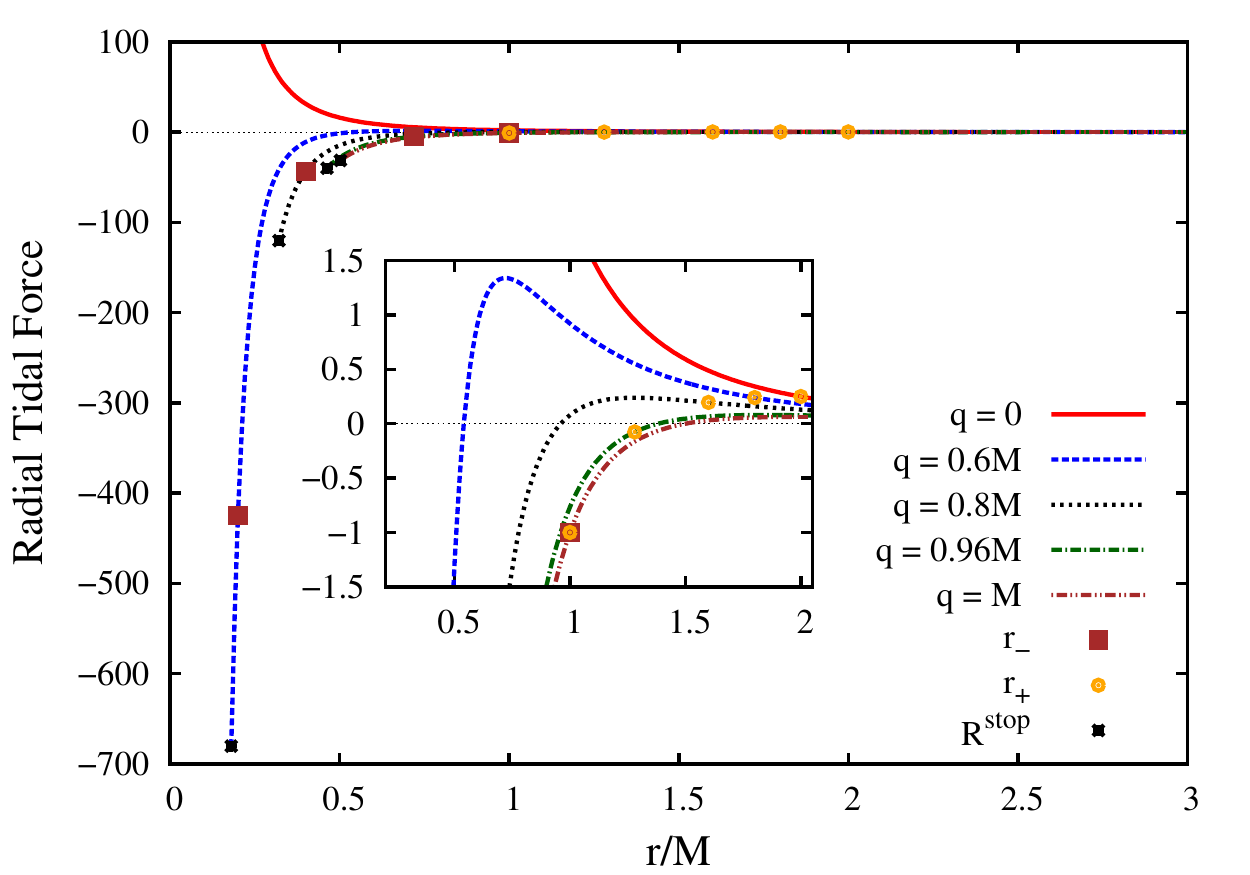}
\caption{Radial tidal force for
Reissner-Nordstr\"om black holes with
different choices of $q/M$
($q=0.6M$, $q=0.8M$, $q=0.96M$, and $q=M$),
as well as for the Schwarzschild black hole
($q=0$).  The positions of $R^{\text{stop}}$, $r_{-}$ and $r_{+}$, are exhibited in each plot. We have chosen $b=100M$. 
%\\blue{Note that the radial tidal force is positive in the region $r>M$.}
}
\label{Radial_Tidal}
\end{figure}

In Fig. \ref{Radial_Tidal} we plot the radial tidal
 force given by Eq.~(\ref{radials}) for Reissner-Nordstr\"om
 black holes for different values of the black hole charge.
 For $q\neq 0$ there is always a local maximum of
 the radial tidal force.
 In the Schwarzschild case ($q=0$) the
 radial tidal force is always positive and diverges (infinite radial stretching)
 as the body approaches the singularity.

%%%%%%%%%%%%%%%%%%%%%%%%%%%%%%%%%%%%%%%
\subsection{Angular Tidal Forces}
\begin{figure}[htpb!]
	\centering
		\includegraphics[width=1.0\columnwidth]{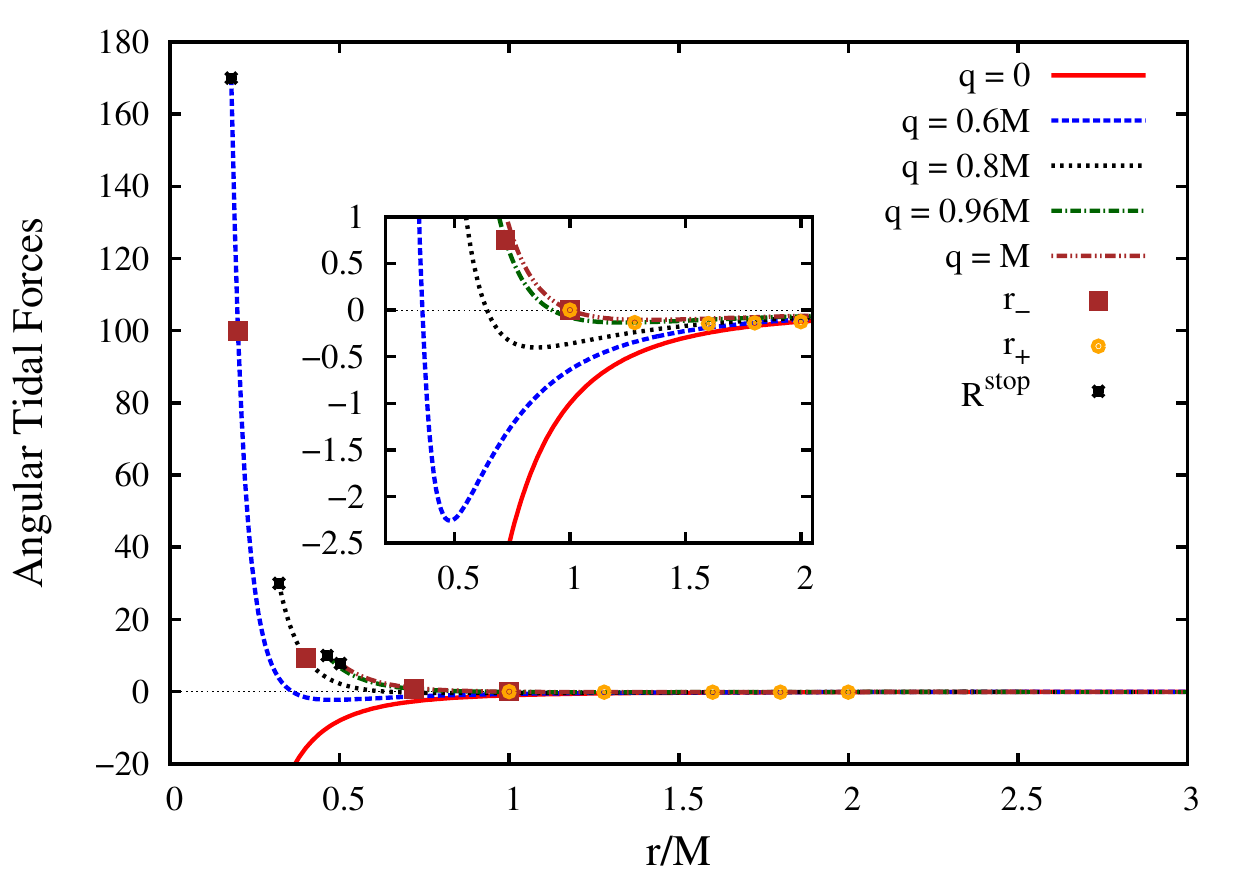}
\caption{Angular tidal forces for
different choices of the charge-to-mass ratio
($q/M=0, \ 0.6, \ 0.8, \ 0.96, \ 1.0$). 
The positions of $R^{\text{stop}}$, $r_{-}$ and $r_{+}$, are exhibited in each plot. We have chosen $b=100M$. 
% \blue{Note that the angular tidal force is negative in the region $r>M$.}
}
\label{Angular_Tidal}
\end{figure}

%%%%%%%%%%%%%%%%%%%%%%%%%%%%%%%%%%%%%%%
We find from Eqs.~\eqref{line} and \eqref{theta} that the angular
 tidal forces vanish at
\beq
\frac{R^{\text{atf}}_{0}}{M} = \frac{q^2}{M^2}
\label{r'}.
\eeq
From Eqs.~(\ref{rh+}),~(\ref{rh-}) and~(\ref{r'}), we obtain
\beq
r_{-} \leq {R^{\text{atf}}_{0}} \leq r_{+} \,\, ,
\label{interval}
\eeq
i.e., the angular tidal forces are zero at some point
between the external and internal horizons, with the equality
in Eq.~(\ref{interval})
holding for the extremely charged black hole.
In Fig.~\ref{Angular_Tidal} we plot the angular tidal
 forces given by Eq.~({\ref{theta}}). 
We note that for $q=0$ (Schwarzschild black hole)
 angular tidal forces are never zero and tend to minus
infinity (infinite angular compressing) as the body approaches the singularity.

\begin{figure}[htpb!]
	\centering
		\includegraphics[width=1.0\columnwidth]{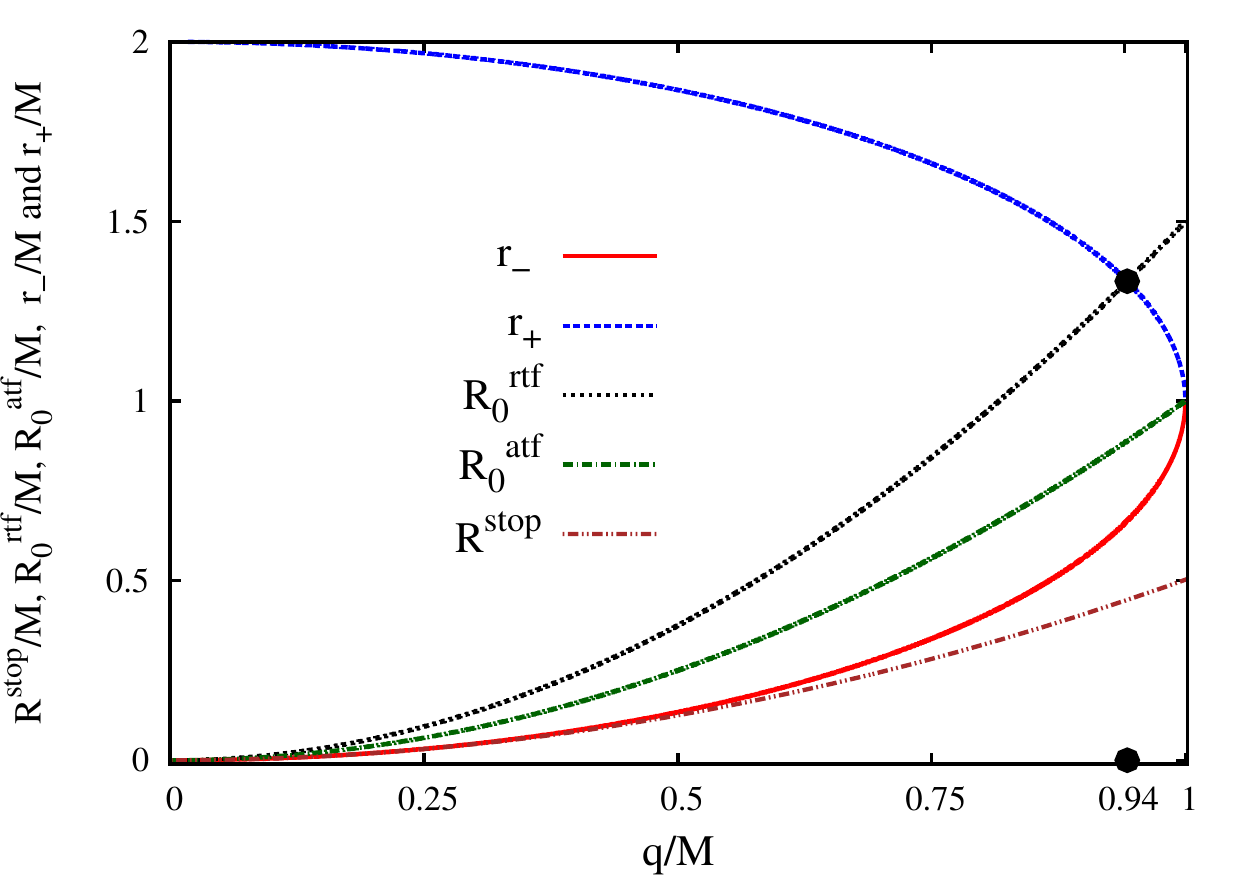}
\caption{$R^{\text{stop}}$, $R_0^{\text{rtf}}$, $R_0^{\text{atf}}$, $r_{-}$, and $r_{+}$,
plotted as functions of $q/M$. The intersection point
of $R_0^{\text{rtf}}$ and $r_{+}$ (black dot) happens at $q/M=2\sqrt{2}/3$,
where the radial tidal force inverts its direction on the event horizon. $R^{\text{stop}}$ is always located inside the internal (Cauchy) horizon. We have chosen $b=100M$.}
\label{rhratf}
\end{figure}

In Fig.~\ref{rhratf} we plot $r_{+}$, $r_{-}$, $R^{\text{stop}}$, $R_0^{\text{rtf}}$ and $R_0^{\text{atf}}$,
given by Eqs.~(\ref{rh+}),~(\ref{rh-}),~(\ref{Rstop}),~(\ref{rtf}) and~(\ref{r'}), respectively,
as functions of $q/M$. We used $b=100M$ to compute $R^{\text{stop}}$.

%%%%%%%%%%%%%%%%%%%%%%%%%%%%%%%%%%%%%%%
\section{Solutions of the Geodesic Deviation Equations in Reissner-Nordstr\"om Spacetimes}
\label{SecVI}
%%%%%%%%%%%%%%%%%%%%%%%%%%%%%%%%%%%%%%%
In this section we solve the geodesic deviation equations~\eqref{radials} and \eqref{theta} and find the geodesic deviation vectors for radially
free-falling geodesics as functions of $r$. As stated in the Introduction, we are considering a 
test body consisting of neutral dust particles infalling radially towards the Reissner-Nordstr\"om black hole.
It is straightforward to convert 
Eqs.~\eqref{radials} and~\eqref{theta} to differential equations in $r$ by using $dr/d\tau = - \sqrt{E^2 - f(r)}$, which results immediately
from Eq.~\eqref{dotr-squared}. Thus, we find
\begin{eqnarray}
\left(E^2-f\right)\eta^{\widehat{1} \prime \prime} 
-\frac{f^{\prime}}{2}\eta^{\widehat{1} \prime}+\frac{f^{\prime \prime}}{2}\eta^{\widehat{1}} & =& 0,
\label{radialr3}\\
\left(E^2-f\right)\eta^{\widehat{i} \prime \prime} 
-\frac{f^{\prime}}{2}\eta^{\widehat{i} \prime}+\frac{f^{\prime}}{2r}\eta^{\widehat{i}} & = & 0.
\label{thetar3}
\end{eqnarray}
The analytical solutions of Eqs.~(\ref{radialr3}) and~(\ref{thetar3}) may be expressed in the following general form, 
for the radial component
\beq
\eta^{\widehat{1}}(r) = \sqrt{E^2-f}\left[C_1+C_2\int \frac{dr}{(E^2 - f)^{3/2}}\right],
\label{gsolr}
\eeq
and for the angular components
\beq
\eta^{\widehat{i}}(r) = \left[C_3 +C_4\int \frac{dr}{r^2(E^2-f)^{1/2}}\right] r,
\label{gsoli}
\eeq
where $C_1$, $C_2$, $C_3$ and $C_4$ are constants of integration. 
Now we specialize to Reissner-Nordstr\"om spacetime.  We consider the geodesic corresponding to a body released from rest at $r=b > r_{+}$.
Then the solutions to the geodesic deviation equations about this geodesic can be written as follows:
\begin{equation}
\eta^{\widehat{1}}(r) = \frac{b^3}{Mb-q^2}\dot{\eta}^{\widehat{1}}(b)\left( E^2-f\right)^{1/2} + \frac{Mb-q^2}{b(2Mb-q^2)}\eta^{\widehat{1}}(b)g(r),
\end{equation}
where
\begin{eqnarray}
g(r) & = & \frac{q^2}{M^2 - (1-E^2)q^2}\left( M - \frac{q^2}{r}\right) + 2r
\nn\\&+& \frac{3}{1-E^2}\left[ r(E^2 -f) + 
\frac{M(E^2-f)^{1/2}}{(1-E^2)^{1/2}}\cos^{-1}\frac{(1-E^2)r - M}{M-q^2/b}\right], \nn\\
\end{eqnarray}
for $ 0 \leq q \leq 1$.

\begin{figure*}[htbp!]
	\centering
\includegraphics[width=1.0\columnwidth]{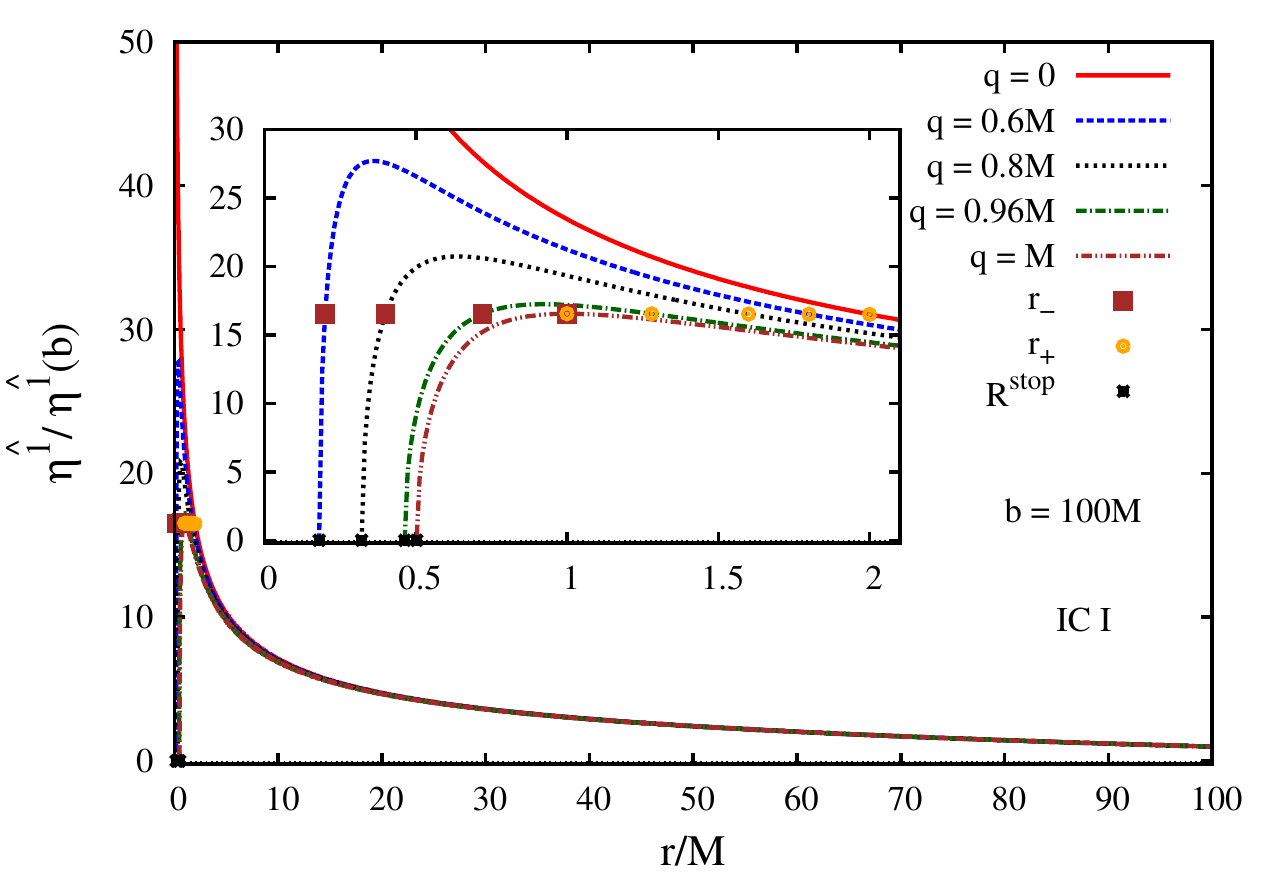}\includegraphics[width=1.0\columnwidth]{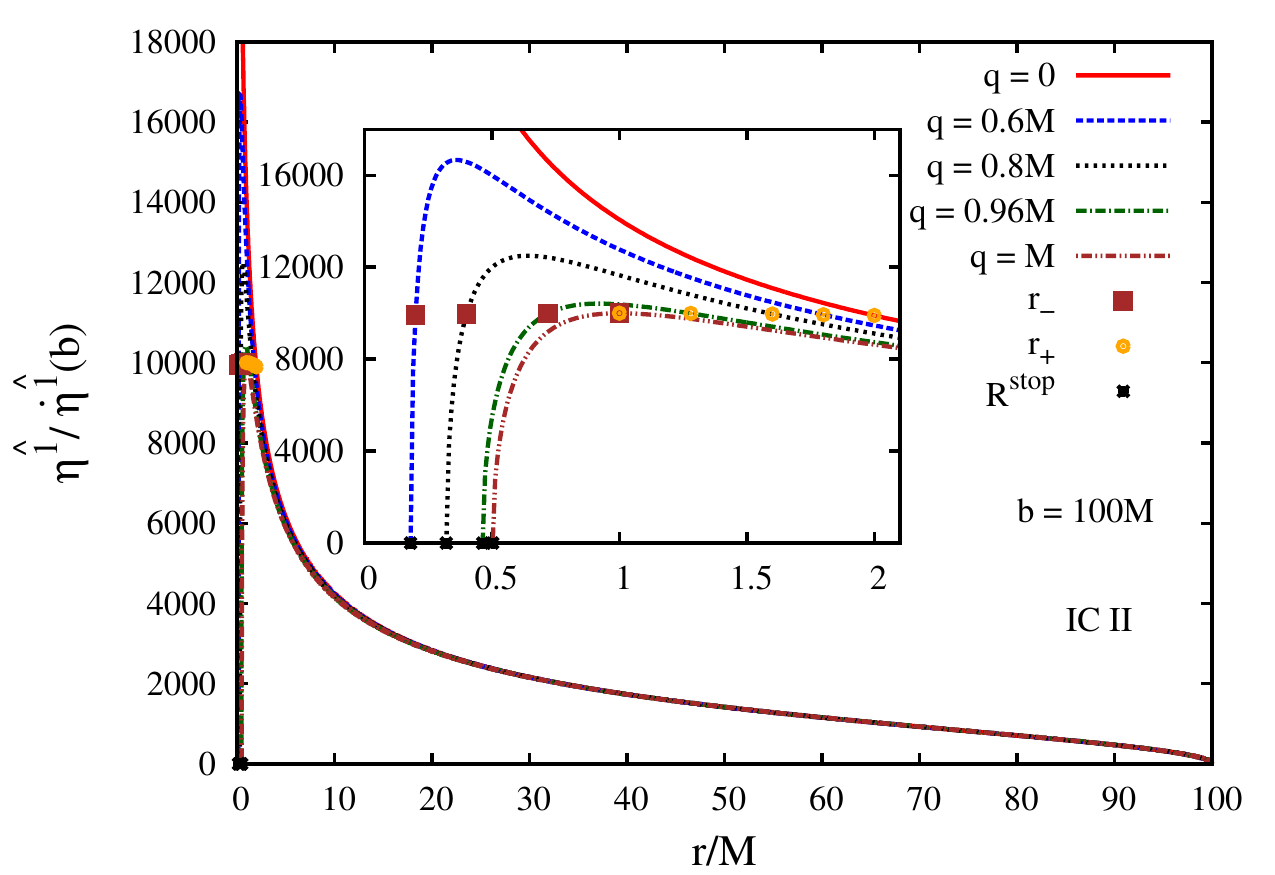}\\
\includegraphics[width=1.0\columnwidth]{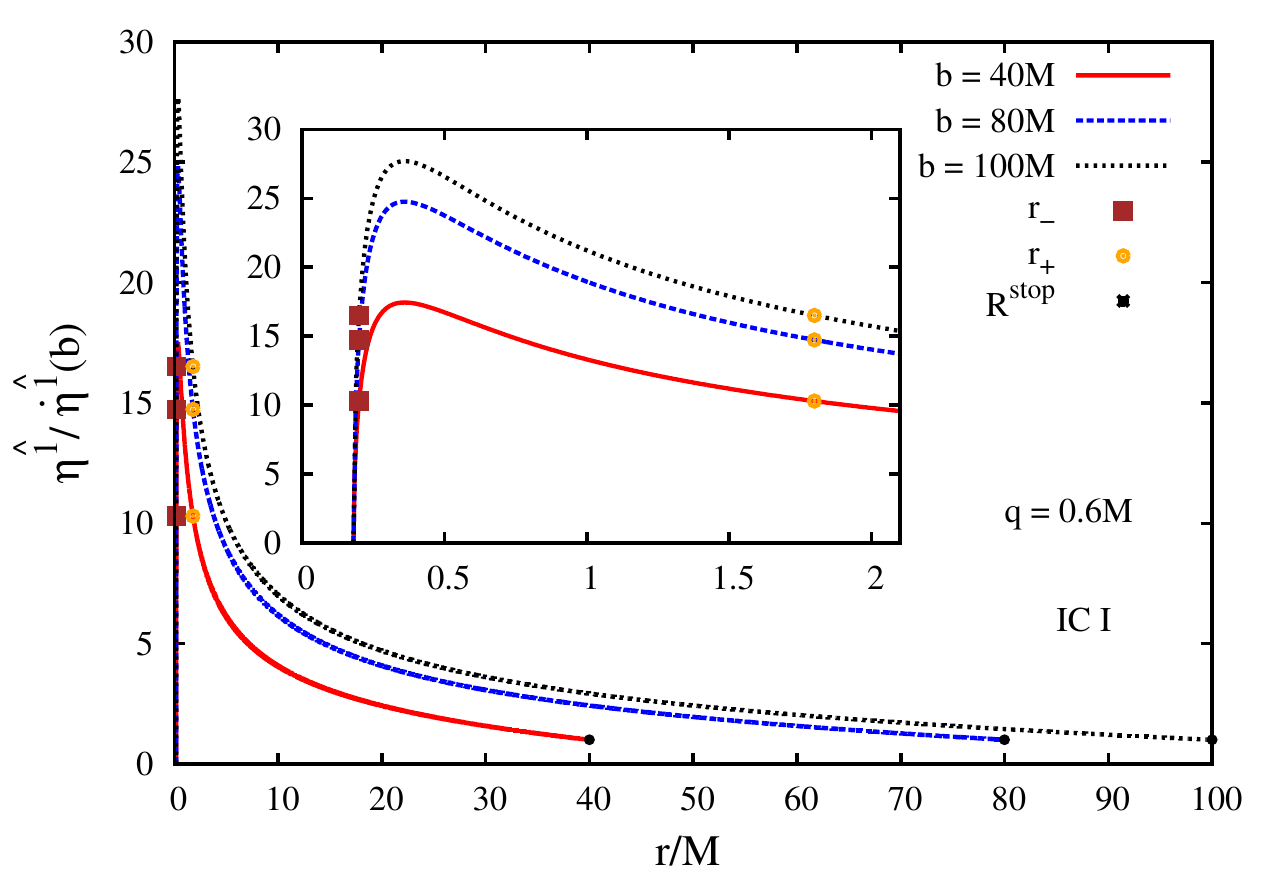}\includegraphics[width=1.0\columnwidth]{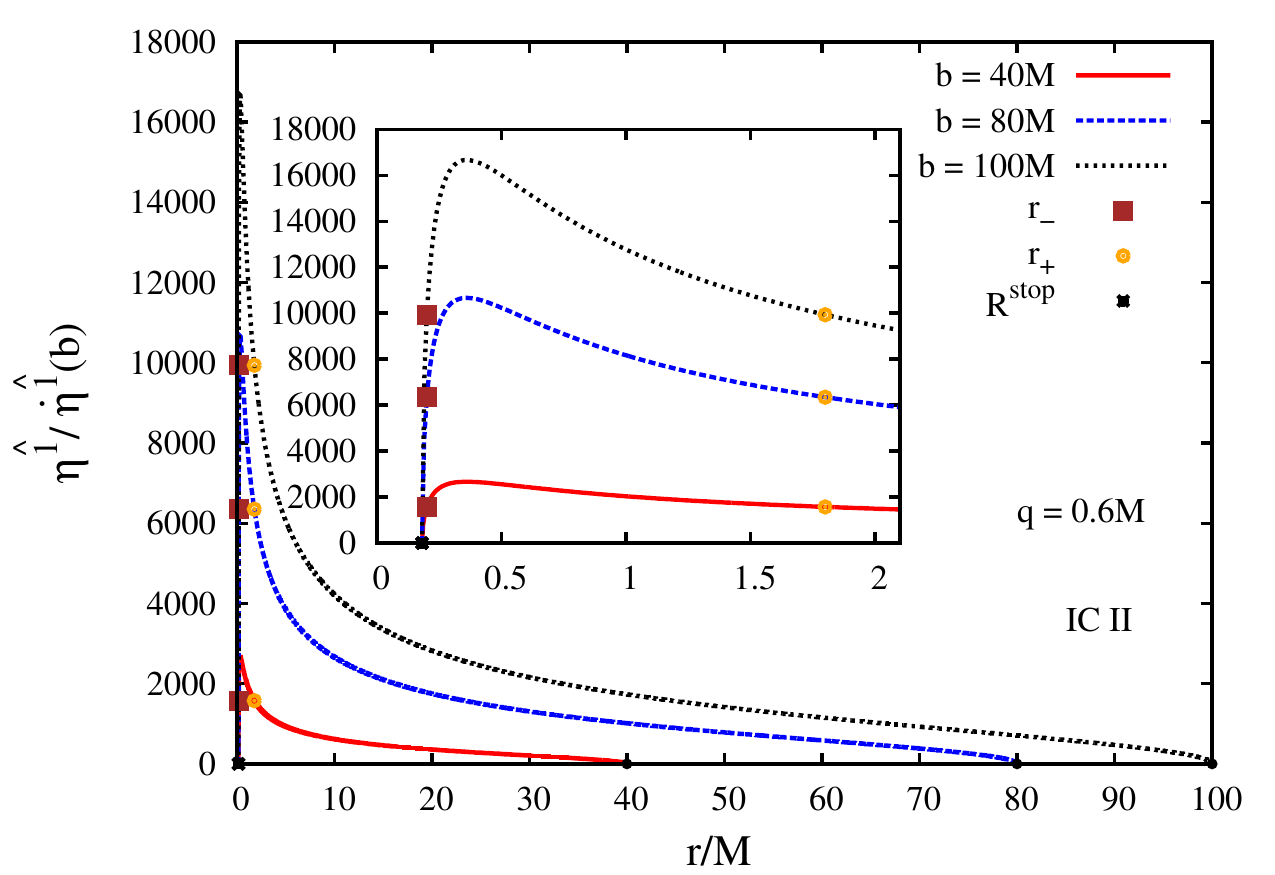}
\caption{Radial components of the geodesic deviation vector for several values of 
charge-to-mass ratio with $b=100M$ and
for $q/M=0.6$ with several values of $b$. The figures on the left correspond to {\rm IC\,I},
whereas the figures on the right correspond to {\rm IC\,II}. The positions of $R^{\text{stop}}$, $r_{-}$ and
$r_{+}$ are exhibited in each plot.}
\label{gdvr}
\end{figure*}

\begin{figure*}[htbp!]
	\centering
\includegraphics[width=1.0\columnwidth]{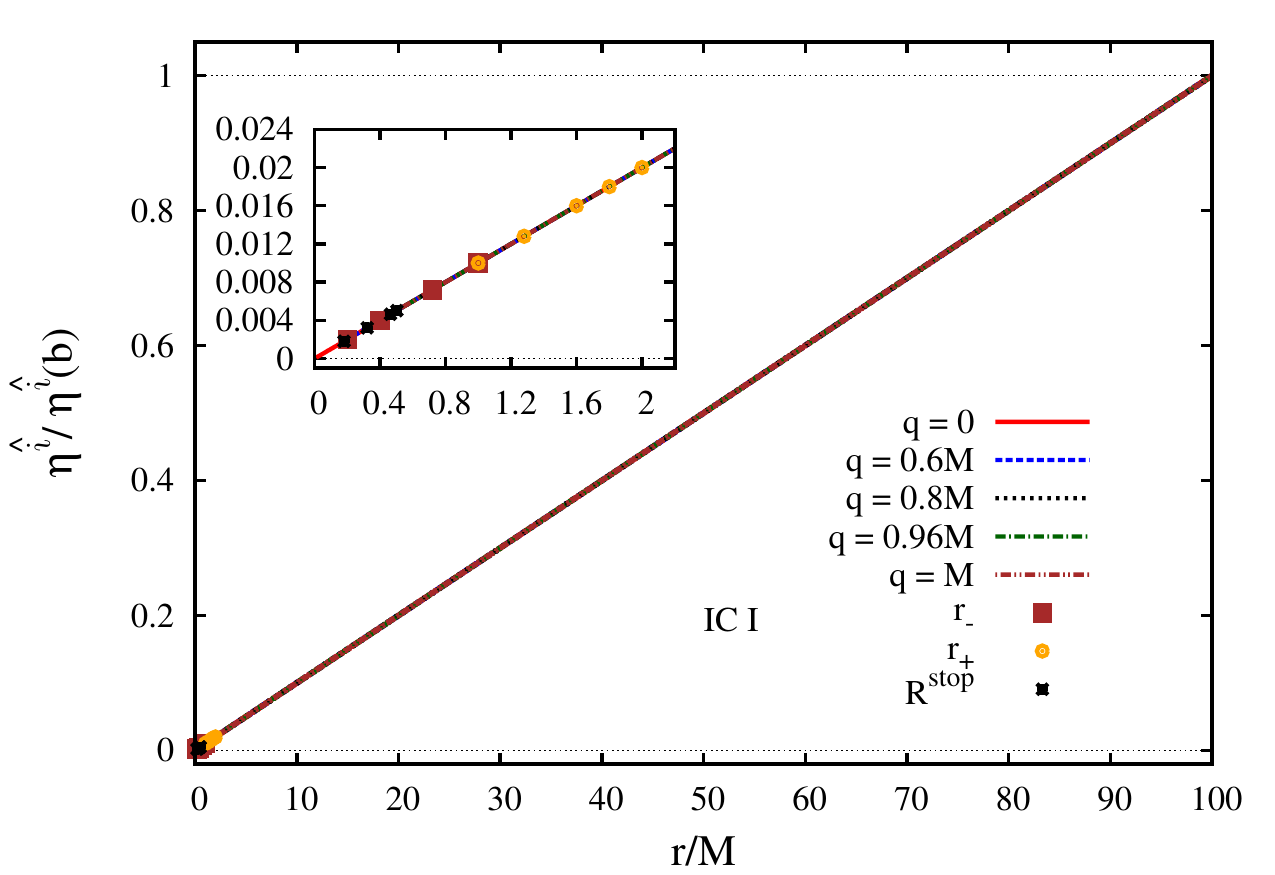}\includegraphics[width=1.0\columnwidth]{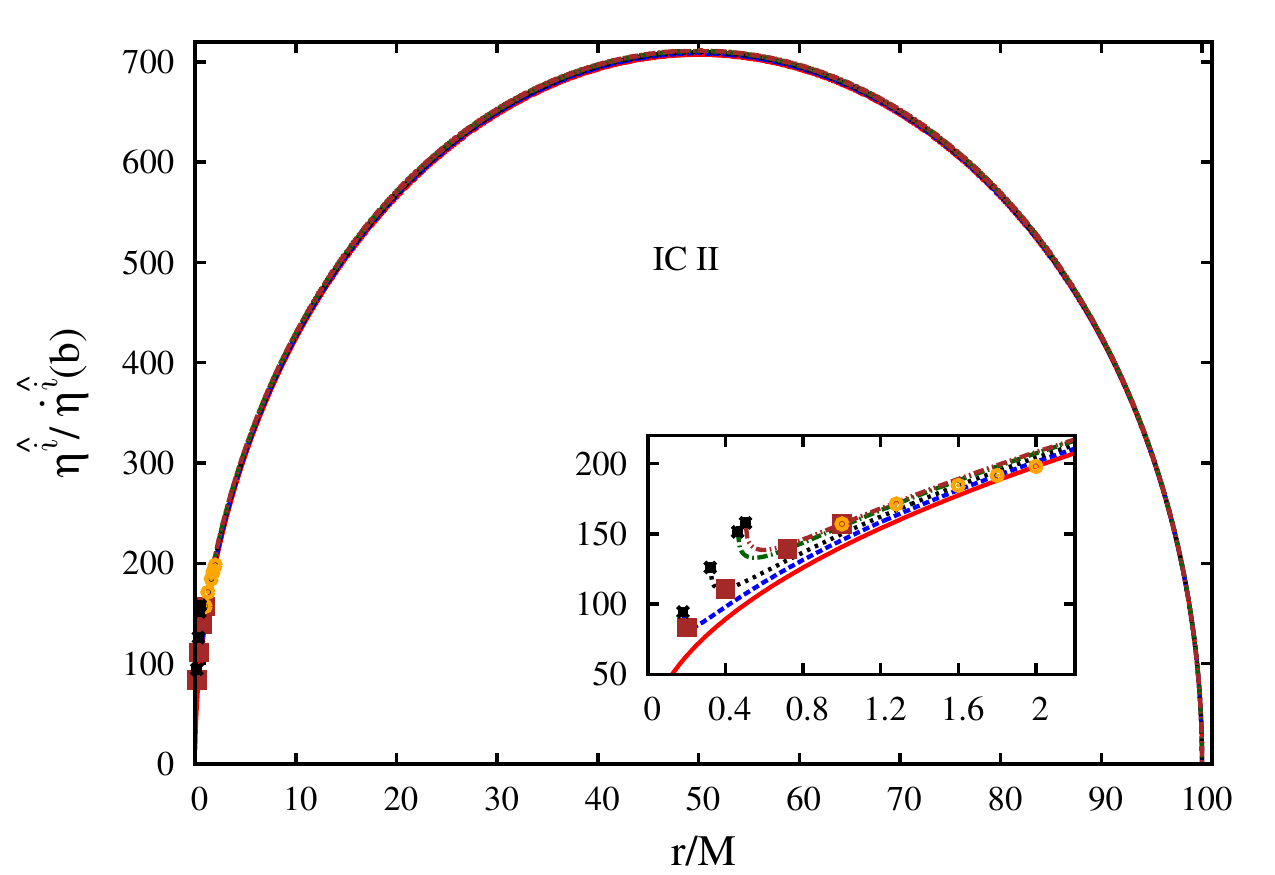}\\
\includegraphics[width=1.0\columnwidth]{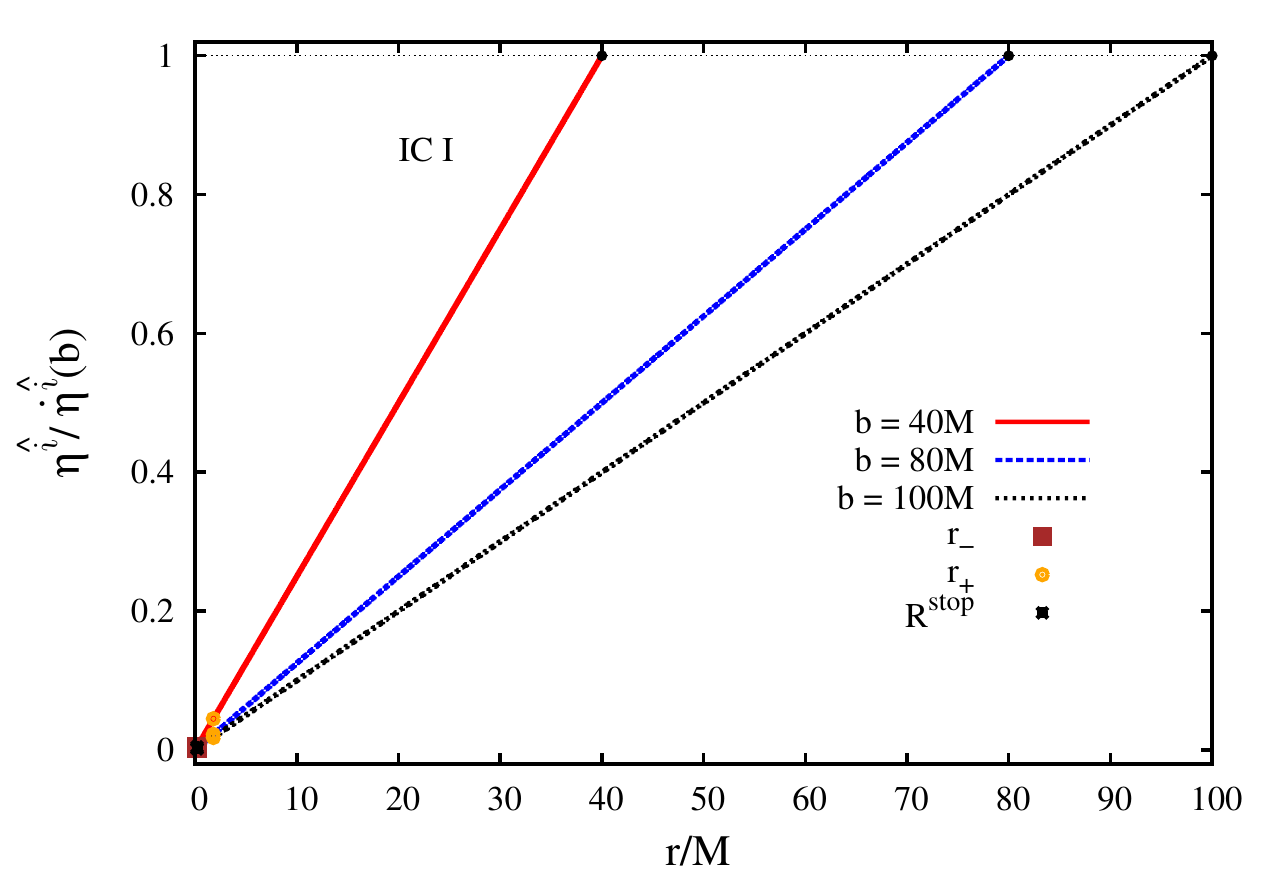}\includegraphics[width=1.0\columnwidth]{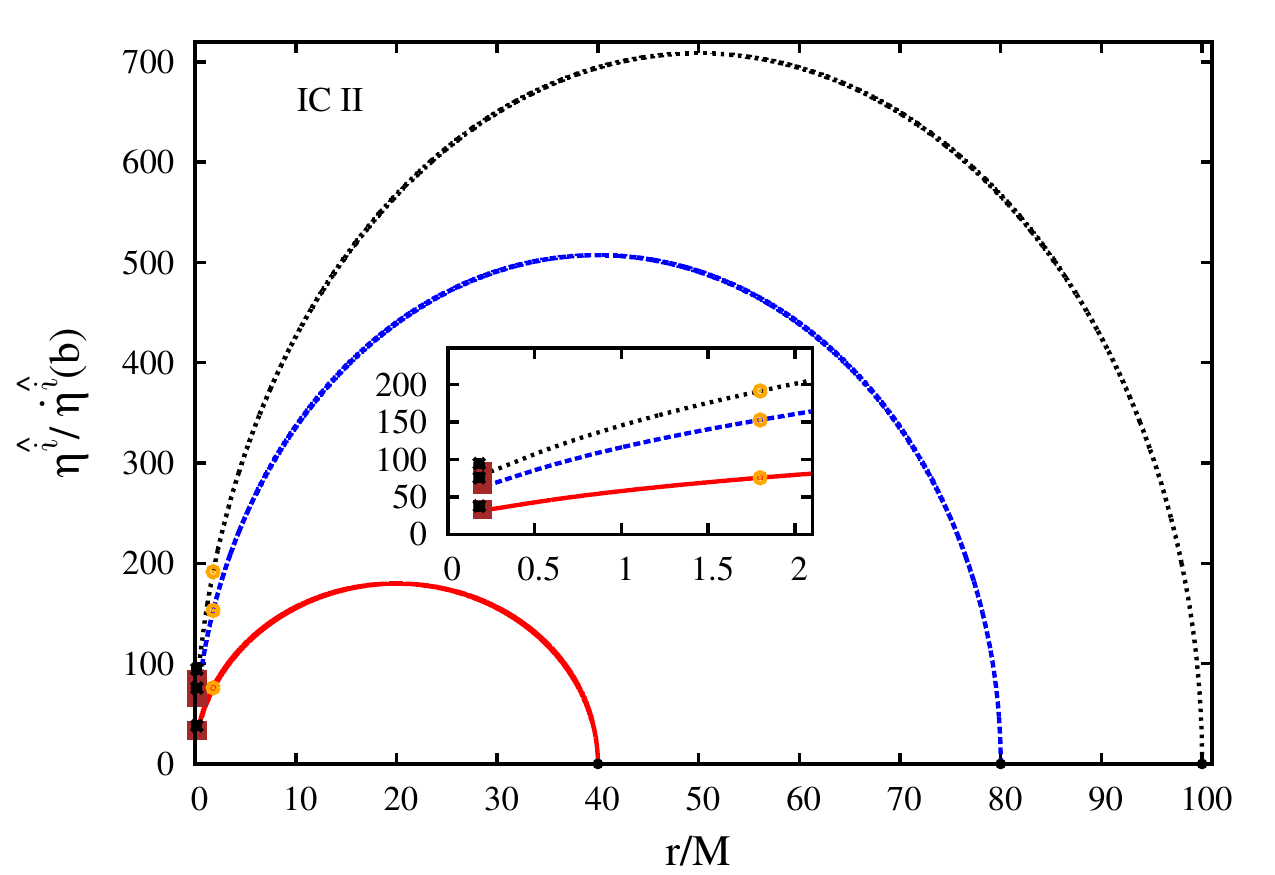}
\caption{Angular components of the geodesic deviation
vector for several values of charge-to-mass ratio with $b=100M$ and for $q/M=0.6$ with several values of $b$.
The positions of $R^{\text{stop}}$, $r_{-}$ and $r_{+}$, are exhibited in each plot.}
\label{gdva}
\end{figure*}

For $q> 0$, the angular components read
\beqn
&&\eta^{\widehat{i}}(r) = \left[ \frac{1}{b}\eta^{\widehat{i}}(b)
+ \frac{b}{q}\dot{\eta}^{\widehat{i}}(b) \cos^{-1}\frac{M-q^2/r}{M-q^2/b}\right]r,\nn\\
\eeqn
and, for $q=0$, they read
\beq
\eta^{\widehat{i}}(r) = \left[ \frac{1}{b}\eta^{\widehat{i}}(b) + \dot{\eta}^{\widehat{i}}(b)
\left[ \frac{2b}{M}\left(\frac{b}{r} - 1\right)\right]^{1/2}\right]r.
\eeq
Here, $\eta^{\widehat{1}}(b)$ and $\eta^{\widehat{i}}(b)$ are the radial and angular components of the initial geodesic deviation vector at $r=b$ and 
$\dot{\eta}^{\widehat{1}}(b)$ and $\dot{\eta}^{\widehat{i}}(b)$ are the corresponding derivatives with respect to the proper time $\tau$.
We plot in Figs.~\ref{gdvr} and \ref{gdva} the
radial and angular components, respectively, 
of the geodesic deviation vector of a body infalling from rest at $r = b$ towards the black hole for different choices of 
the black hole charge. Here we consider two different initial conditions, {\rm IC\,I} and {\rm IC\,II}, for the 
radial and angular components of the geodesic deviation vector at $r=b$. For the first type, {\rm IC\,I}, we choose $\eta^{\widehat{1}}>0$, $\eta^{\widehat{i}}>0$, $\dot{\eta}^{\widehat{1}}=0$, and $\dot{\eta}^{\widehat{i}}=0$, at $r = b$. For the second type, {\rm IC\,II}, we choose  $\eta^{\widehat{1}}=0$, $\eta^{\widehat{i}}=0$, $\dot{\eta}^{\widehat{1}}>0$, and 
$\dot{\eta}^{\widehat{i}}>0$, at $r = b$. The condition {\rm IC\,I} corresponds to 
releasing a body consisting of dust at rest with no internal motion. The condition {\rm IC\,II}, on the other hand, corresponds to letting
such a body `explode' from a point at $r=b$.
The behavior of the geodesic deviation vector is almost identical for different values of $q$ until $r$ becomes 
of the same order as the horizon radius. This is as expected because
for large $r$ the spacetime looks similar for all values of $q$.   It can be seen from Fig.~\ref{gdvr} that, for
{\rm IC\,II} (figures on the right), during the infall
from $r=b$ to $r_{+}$, the radial
 component always increases.
If $q \neq 0$, while the body falls from the outer horizon ($r=r_{+}$) to the inner horizon ($r_{-}$),
 the radial component of the geodesic deviation vector keeps increasing, reaches a maximum, and then starts decreasing until it reaches $r_{-}$.
While the body falls in the region 
between $r_{-}$ and $R^{\text{stop}}$, the radial component of the geodesic deviation vector keeps
decreasing, finally shrinking to zero at $R^{\text{stop}}$. (Recall, however, that, since the 
inner (Cauchy) horizon is unstable, the region between $r=r_{-}$ and $R^{\text{stop}}$ is unphysical.) 
The radial component of the geodesic deviation vector 
with the initial condition {\rm IC\,I} 
(shown on the left of Fig.~\ref{gdvr})
behaves similarly to {\rm IC\,II}, 
except that in the former case it becomes zero at some $r$
satisfying $R^{\text{stop}} < r < r_{-}$~\footnote{For the radial component of the geodesic deviation vector with the initial condition {\rm IC\,I}, we have~$\eta^{\widehat{1}}(r)|_{r=R^{\text{stop}}}\simeq-10^{-6}$ (i.e., it is zero at some value of $r$ satisfying $R^{\text{stop}} < r < r_{-}$), while for initial condition {\rm IC\,II}, we have~$\eta^{\widehat{1}}(r)|_{r=R^{\text{stop}}}=0$.}.

With the initial conditions {\rm IC\,II}, it can be seen from Fig.~\ref{gdva} (figures on the right) that the angular 
components
increase in the beginning but start decreasing around $r=b/2$, reflecting the compressing nature of 
the angular tidal force. 
They then start increasing at some point before the body reaches $r=R^{\text{stop}}$ if $q \neq 0$ because of the change in the sign of the angular components
of the tidal force.  
If the initial conditions are {\rm IC\,I}, the angular components of the geodesic deviation 
vector decrease linearly in $r$, as shown in the figures on the left 
 of Fig.~\ref{gdva}.  This is expected because all geodesics with no angular velocity trace out straight radial lines.  

We also can see from these figures
that for Schwarzschild spacetime ($q=0$)
the radial component of the geodesic deviation vector becomes
infinite at the singularity $r=0$ whereas the angular components vanish there,
as expected~\cite{MTW, gad}.

%%%%%%%%%%%%%%%%%%%%%%%%%%%%%%%%%%%%%%%
\section{Conclusion}
\label{sectionIV}
%%%%%%%%%%%%%%%%%%%%%%%%%%%%%%%%%%%%%%%
In this paper we investigated tidal forces in
 Reissner-Nordstr\"om spacetimes,
which depend on the charge-to-mass ratio of the black hole. For certain
values of these parameters, the radial tidal force can change from 
stretching to compressing outside the event horizon.
We also noted that angular tidal forces can only be zero
 between the two horizons of the charged black hole.

We pointed out that the tidal forces
in Schwarzschild and Reissner-Nordstr\"om
 spacetimes can be quite different close to the black hole.
In Schwarzschild spacetime the tidal forces always cause stretching 
 in the radial direction and compression in the angular
 directions whereas in Reissner-Nordstr\"om spacetime they may cause either stretching or
 compression in any direction, depending on the
 charge-to-mass ratio of the Reissner-Nordstr\"om black hole and the radial coordinate.

We also noted that the geodesic deviation equations about a radially free-falling geodesic
can be solved analytically.
In particular, we examined the behavior of the geodesic deviation vector for such a geodesic
under the influence of tidal forces created by
static charged black holes.
We noted that the behavior of the geodesic deviation vector
is qualitatively the same for both Schwarzschild and Reissner-Nordstr\"om
black holes away from the horizon.
However, its behavior is considerably different for Schwarzschild and
Reissner-Nordstr\"om black holes inside
the event horizon, as we
showed in Sec. \ref{SecVI}.
For instance, for charged black holes, at its closest approach
to the singularity, the radial component of the geodesic deviation vector becomes zero for a certain initial condition
though its angular components remain finite there (though this point is beyond the Cauchy horizon and, hence,
in an unphysical region).  In contrast, for the uncharged black hole
the radial component of the geodesic deviation vector increases all the way to the singularity
whereas the angular components shrink to zero at the singularity.

%%%%%%%%%%%%%%%%%%%%%%%%%%%%%%%%%%%%%%%
\section*{Acknowledgments}
 The authors would like to thank Erberson R. Pinheiro and
Jo\~ao V. N. de Ara\'ujo for their contributions in the early stages of this work. The authors acknowledge
Conselho Nacional de Desenvolvimento
Cient\'\i fico e Tecnol\'ogico,
Coordena\c{c}\~ao 
de Aperfei\c{c}oamento de Pessoal
de N\'\i vel Superior and Funda\c{c}\~ao Amaz\^onia de Amparo a Estudos e Pesquisas do Par\'a
for partial financial support. AH also
 acknowledges partial support from the Abdus Salam
 International Centre for Theoretical Physics through Visiting Scholar/Consultant Programme. AH thanks the
 Universidade Federal do Par\'a in Bel\'em for kind hospitality.

%%%%%%%%%%%%%%%%%%%%%%%%%%%%%%%%%%%%%%%

%%%%%%%%%%%%%%%%%%%%%%%%%%%%%%%%%%%%%%%%%%%%%%%%%%%%%%%%%%%%%%%%%%%%%%%%%%%%%%%%%%
\end{document}